\documentclass[conference]{IEEEtran}

\usepackage{amsmath,amssymb,amsfonts,amsthm}
\usepackage{url}
\usepackage{cite}

\usepackage{xcolor}
\usepackage{tabularx}
\usepackage{xspace}
\usepackage[caption=false,font=footnotesize]{subfig}

\usepackage{tikz}
\usetikzlibrary{decorations.pathreplacing}
\usetikzlibrary{calc}
\usetikzlibrary{patterns}
\tikzset{font=\small}

\usepackage{listings}
\makeatletter
\def\lst@makecaption{%
  \def\@captype{table}%
  \@makecaption
}
\makeatother

\usepackage[shortcuts]{extdash}
\usepackage[acronym]{glossaries}
\newacronym{bpm}{BPM}{Business Process Management}
\newacronym{bpmn}{BPMN}{Business Process Model and Notation}
\newacronym{mno}{MNO}{Mobile Network Operator}
\newacronym[plural=BPMS,firstplural=Business Process Management Systems (BPMS)]{bpms}{BPMS}{Business Process Management System}
\newacronym{ntp}{NTP}{Network Time Protocol}
\newacronym{utc}{UTC}{Coordinated Universal Time}

\newtheorem{definition}{Definition}

\usepackage[absolute]{textpos}
\newcommand{\ieeecopyright}{
  \begin{textblock}{7}(0.7,10.15)
    \noindent\fbox{\parbox{\textwidth}{
    {\small
    {\itshape Accepted article at the 24th IEEE EDOC 2020 conference.}\\
    {\bfseries © 2020 IEEE.} Personal use of this material is permitted.  Permission from IEEE must be obtained for all other uses, in any current or future media, including reprinting/republishing this material for advertising or promotional purposes, creating new collective works, for resale or redistribution to servers or lists, or reuse of any copyrighted component of this work in other works.}
    }}
  \end{textblock}
}

\begin{document}

\title{Time in Blockchain-Based Process Execution}

\author{
\IEEEauthorblockN{Jan Ladleif and Mathias Weske}
\IEEEauthorblockA{Hasso Plattner Institute\\
University of Potsdam\\
Potsdam, Germany\\
\{jan.ladleif, mathias.weske\}@hpi.uni-potsdam.de}
}

\maketitle

\begin{abstract}
The traceable execution of business processes and choreographies using smart contracts is one prominent application of blockchain technology in \gls{bpm}.
Existing approaches support a large set of patterns, modeling languages, and blockchain architectures, which cover a wide range of practical scenarios.
However, they largely neglect the important aspect of \emph{time}, a crucial part of process and choreography models manifested in deadlines, delays, and other temporal constraints.
We argue that this deficit is due to inherent limitations of smart contracts---in particular the absence of a natural notion of measuring time---on popular blockchain platforms used in research and practice.
We introduce a set of time measures available on blockchain platforms to alleviate these issues, and systematically compare their properties.
We also give hints as to their suitability for facilitating various temporal constraints commonly found in process models.
\end{abstract}

\begin{IEEEkeywords}
executable processes, blockchain, smart contracts, temporal constraints
\end{IEEEkeywords}

\glsresetall

\newcommand{\mtimestamp}{M_{\text{BT}}}
\newcommand{\mnumber}{M_{\text{BN}}}
\newcommand{\mparam}{M_{\text{PA}}}
\newcommand{\mpull}{M_{\text{RO}}}
\newcommand{\mpush}{M_{\text{SO}}}
\newcommand{\pull}{request\-/response\xspace}
\newcommand{\Pull}{Request\-/response\xspace}
\newcommand{\push}{storage\xspace}
\newcommand{\Push}{Storage\xspace}

\ieeecopyright

\section{Introduction}
\label{sec:introduction}

In \gls{bpm}, process and choreography models are used to specify processes within and between organizations, respectively~\cite{weske2019business}.
While models often serve documentation purposes only, they are also used as source artifacts in model\-/driven process execution on different target platforms.
One such target platform are blockchain\-/based smart contracts, which have been shown to enable transparent, tamper\-/proof, and auditable process execution which is especially valuable for inter\-/organizational communication between untrusted partners~\cite{weber2016untrusted}.

Sophisticated approaches have emerged, allowing users to employ a large bandwidth of process modeling tools~\cite{tran2018lorikeet,lopez2019caterpillar,ladleif2019modeling,klinger2020cross}.
They have a drawback in common, however; they do not properly take into account time, which is of paramount importance in process models in general~\cite{eder1999time} and in contractual agreements in particular~\cite{ladleif2019unifying}.
Deadlines have to be kept and timeouts have to be observed closely for partners to achieve a common business goal, and process modeling languages naturally provide a multitude of elements to express such temporal constraints~\cite{cheikhrouhou2015temporal}.

This deficiency is ostensibly due to the peculiar properties and mining protocols of commonly used blockchain platforms.
Smart contracts are entirely transaction\-/driven and executed within a closed\-/world environment without access to global timing information.
As such, a smart contract executing certain parts of the model can often not reliably decide whether a temporal constraint is violated or not.

In this paper, we want to alleviate these issues in a four\-/step approach.
First, (i) an abstract formal model of blockchain is introduced, based on which (ii) a set of time measures which can be used within the narrow confines of a smart contract is developed.
We then (iii) critically compare properties of the measures, such as accuracy and trust, before (iv) discussing how they can be used to implement generic temporal constraints found in process models.
By doing so, we hope to give structured insights into the techniques which can be used by blockchain\-/based process execution approaches, and provide guidelines on how to chose the right measure for a given approach.

\begin{figure*}
  \centering
  \includegraphics[scale=.61]{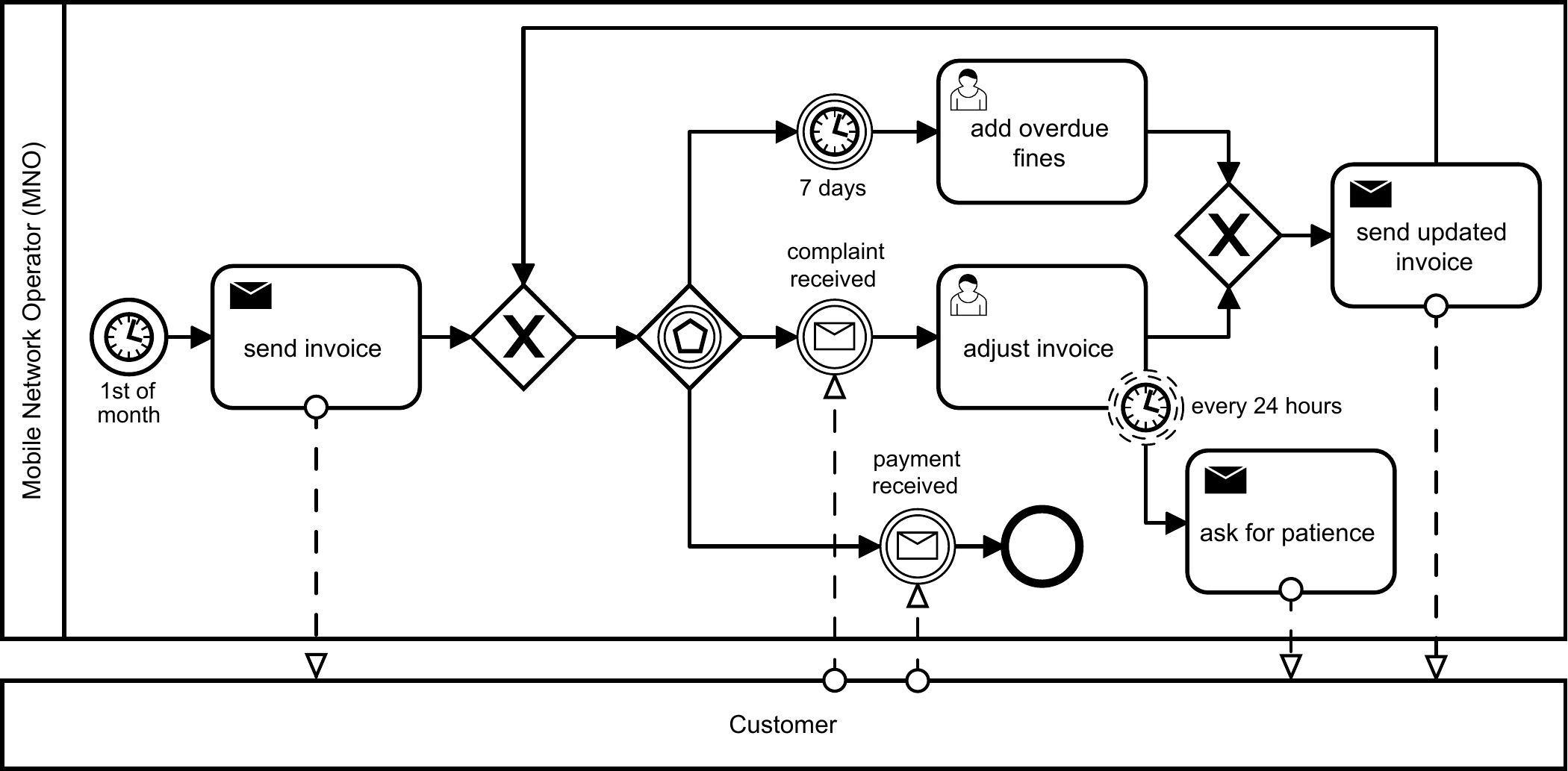}
  \caption{Running example of a simple choreography between an \gls{mno} and its customer for sending and handling an invoice}
  \label{fig:example}
\end{figure*}

As a motivating example throughout the paper, we will consider a simplified invoicing choreography between a \gls{mno} and one of their customers.
Figure~\ref{fig:example} shows a view of the orchestration process of the \gls{mno} modeled using the de\-/facto industry standard \gls{bpmn}~\cite{omg2013bpmn}.
It starts on the 1\textsuperscript{st} of each month with the sending of an invoice.
The customer may pay directly, or be reminded every seven days incurring additional overdue fines.
Also, they may file a complaint with the \gls{mno}, e.g., because of an incorrect charge on the invoice.
The invoice is then manually adjusted, and a message asking for patience is sent regularly every 24 hours to keep the customer informed.
The example illustrates that temporal constraints are pervasive in business processes, and blockchain\-/based process execution approaches necessarily need to support such scenarios in order to gain widespread acceptance.

The paper is structured as follows.
Preliminary knowledge and a detailed problem statement are conveyed in Sect.~\ref{sec:background}.
A formalization of the abstract structure of blockchains is provided in Sect.~\ref{sec:formalization}, which is used to define and assess a set of time measures available in blockchain\-/based process execution in Sect.~\ref{sec:measures}.
We gauge whether the time measures are suitable for implementing temporal constraints in Sect.~\ref{sec:semantics}, before critically discussing and evaluating the results in Sect.~\ref{sec:discussion}.
We review related work based on our findings in Sect.~\ref{sec:related-work}, and finally conclude in Sect.~\ref{sec:conclusion}.

\section{Background}
\label{sec:background}

This section establishes preliminary knowledge pertaining to temporal constraints in business processes and the underlying techniques and restrictions of blockchain technology as well as smart contracts.
The state of the art of blockchain\-/based process execution is then assessed on that basis to further narrow down the problem statement at hand.

\subsection{Temporal Constraints in Business Process Management}
\label{subsec:time-in-bpm}
Temporal constraints are ubiquitous in business processes and choreographies and have been an active field of research for decades~\cite{eder1999time}.
They can be broadly divided into three categories: \emph{intra\-/activity}, e.g., durations of atomic tasks; \emph{inter\-/activity}, e.g., delays between activities and events; and \emph{inter\-/process} temporal constraints, i.e., deadlines between partners in a choreography~\cite{cheikhrouhou2015temporal}.

The running example in Fig.~\ref{fig:example}, for instance, contains the latter two categories of constraints, modeled using timer events discernible by the clock symbol.
They have different purposes, like initially starting a process instance at the \emph{``1st of [the] month''} which then indirectly triggers the customer's local orchestration process (inter\-/process), or expressing delays such as \emph{``7 days''} in the process flow (inter\-/activity).

In any case, timer events are further refined from their natural language label by a formal specification.
\gls{bpmn} largely follows a generic distinction between absolute and relative temporal constraints~\cite{cheikhrouhou2015temporal}, and allows for three different notions of timer events following the ISO\-/8601 standard (see Tab.~\ref{tab:definitions})~\cite[p.~273]{omg2013bpmn}.
\emph{Dates} are used to wait for a specific, absolute point in time; \emph{durations} are used to delay process execution or represent timeouts relative to the event's enablement; and \emph{cycles} are used to model actions that are taken in a recurring fashion.
The latter may start at an absolute point in time or relative to the event's enablement.

\begin{table}[b]
\renewcommand{\arraystretch}{1.1}
\caption{Timer Event Definitions in \gls{bpmn} in Accordance to ISO\-/8601}
\label{tab:definitions}
\centering
\begin{tabular}{l|l}\hline
\emph{Name} &
\emph{Example}
\\\hline\hline

Date &
\verb|2020-12-24T12:00:00Z|
\\ &
(noon on Dec. 24\textsuperscript{th}, 2020, UTC)
\\\hline

Duration &
\verb|P7D|
\\ &
(7 days)
\\\hline

Cycle (abs.) &
\verb|R/2020-01-01/P1M|
\\ &
(every 1\textsuperscript{st} of the month from 2020)
\\\hline

Cycle (rel.) &
\verb|R7/PT24H|
\\ &
(every 24 hours for 7 days)
\\\hline

\end{tabular}
\end{table}

Another more indirect temporal constraint is also found in the example, a common pattern known as \emph{deferred choice}~\cite{russell2016workflow}.
It is used in situations in which the process waits for one of several events to happen, and picks the one ``first triggered''~\cite[p.~437]{omg2013bpmn} for the final execution.
In \gls{bpmn}, the pattern is modeled using event\-/based gateways, diamond\-/shaped elements containing two circles and a pentagon, which create a kind of temporal race between the following events.

\subsection{Blockchain Technology and Smart Contracts}
\label{subsec:blockchain}
Graduating from its original motivation in the finance domain~\cite{nakamoto2008bitcoin}, blockchain technology has since emerged as a core platform for secure trustless applications~\cite{xu2017taxonomy}.
The blockchain does not only store a list of account balances anymore, but also application logic in the form of executable code and its associated state, which is generally called a \emph{smart contract}.

Smart contracts are not continuously running but triggered externally using \emph{transactions} propagated in the network.
Transactions and their effects are only applied to the blockchain's overall state during \emph{mining}, a procedure in which information is appended to the blockchain as a new block.
In order to generate a block, a network participant---the \emph{miner}---bundles a set of pending transactions, assigns them an order, and sequentially applies them, altering account balances and executing smart contract code.
Once finished, the block is forwarded to the other nodes in the network to append it to their local versions of the blockchain.

All of these steps are secured via cryptographic schemes, consensus protocols, and incentivization measures depending on the blockchain implementation, with numerous complex extensions being constantly explored and proposed.
Ultimately, this interplay of techniques results in the strong security guarantees of blockchain technology such as integrity, transparency, and immutability, despite potentially distrustful participants in the network~\cite{xu2017taxonomy}.

\subsection{Blockchain Oracles}
\label{subsec:oracles}
\begin{figure}
  \centering
  \subfloat[\Push oracle]{\includegraphics[width=.8\linewidth]{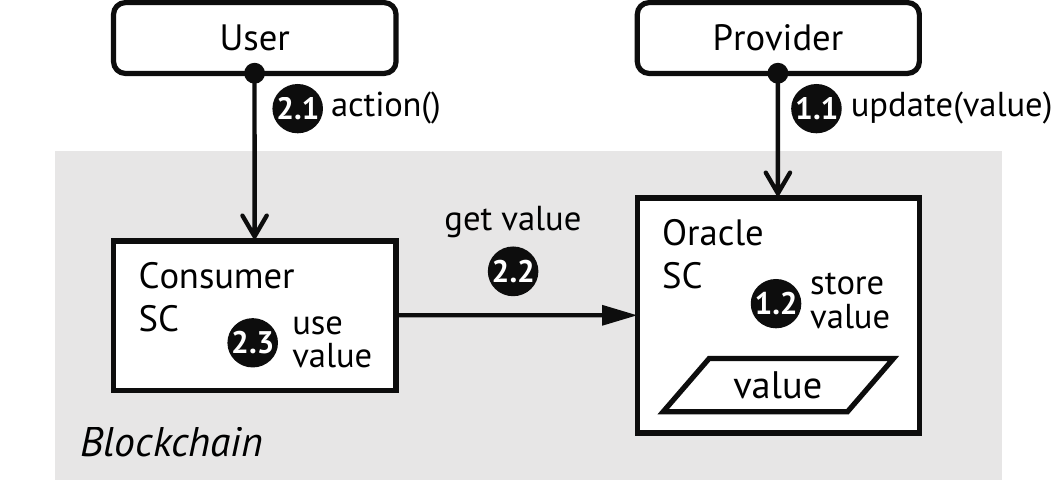}
  \label{fig:push-oracle}}
  \hfil
  \subfloat[\Pull oracle]{\includegraphics[width=.8\linewidth]{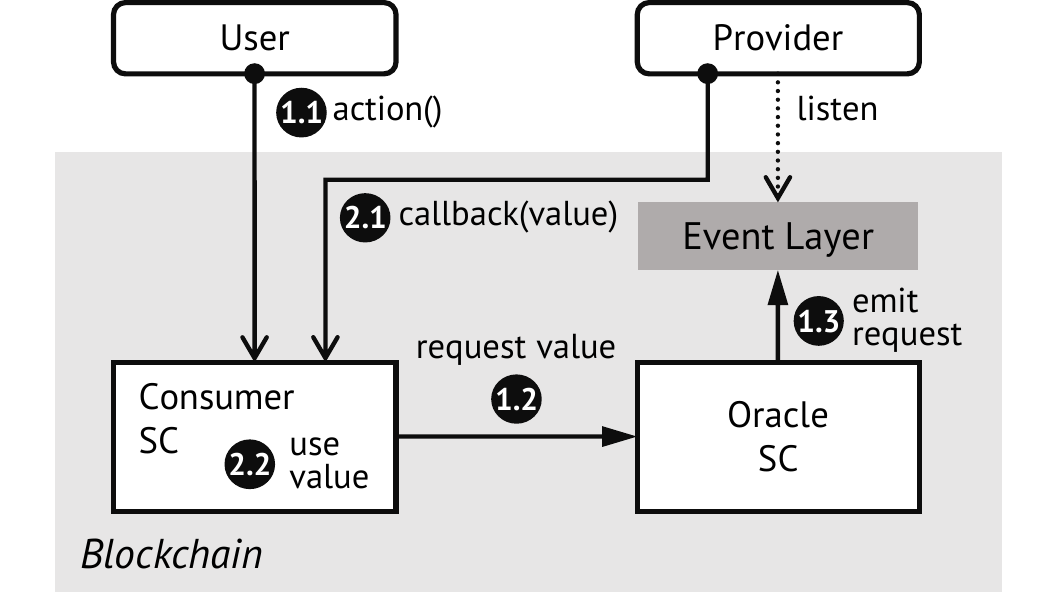}
  \label{fig:pull-oracle}}
  \caption{Schematic usage of oracles}
  \label{fig:oracles}
\end{figure}

A direct consequence of the blockchain's favorable properties is a strict closed\-/world assumption, i.e., data external to the blockchain which could be altered or go missing can not be accessed from within smart contracts for integrity reasons.
Rather, the blockchain needs to be entirely self\-/contained for later validation.
Common patterns to somewhat mitigate this restriction use \emph{oracles}~\cite{xu2018pattern}.
Oracles are operated by third\-/party providers and supply external data to the blockchain.
They generally work using one of two schemes~\cite{albreiki2020trustworthy}:

The data may be stored directly on the blockchain for immediate read access, which we will call \emph{\push oracle}.
For this purpose, an oracle provider maintains a publicly known smart contract (see Fig.~\ref{fig:push-oracle}) and regularly updates it via dedicated transactions (steps 1.1, 1.2).
The data can then be synchronously read by consumer smart contracts (steps 2.1--2.3).
One example for such a service in practice is OrFeed\footnote{\url{https://www.orfeed.org/docs/}, 2020-04-15}, which provides cryptocurrency exchange rates on Ethereum using a similar mechanism.

Alternatively, data may be asynchronously requested from within the blockchain using \emph{\pull oracles} (see Fig.~\ref{fig:pull-oracle}).
Here, the oracle provider operates a publicly known smart contract acting as a proxy.
Data requests directed at this smart contract (steps 1.1, 1.2) are emitted using the blockchain's event layer, which the provider actively listens to (step 1.3).
The query is performed off\-/chain, and the provider sends a transaction with the result of the query back to the original requester (steps 2.1, 2.2).
\Pull oracles might provide general\-/purpose services, like Provable\footnote{\url{http://provable.xyz/}, 2020-03-19}, or more domain\-/specific ones such as Ethereum Alarm Clock\footnote{\url{https://www.ethereum-alarm-clock.com/}, 2020-03-19}.

\subsection{Blockchain\-/Based Execution of Business Process}
\label{subsec:blockchain-based-execution}
The potential of blockchain technology in the area of \gls{bpm} has been widely acknowledged~\cite{mendling2017opportunities}.
One field of particular interest involves the blockchain\-/based execution of business processes, especially in the context of contractual interactions between distrustful partners benefiting from the platform's strong security guarantees~\cite{weber2016untrusted}.

Approaches like Caterpillar~\cite{lopez2019caterpillar} or Lorikeet~\cite{tran2018lorikeet} collectively provide support for a large set of modeling languages and elements and can be roughly categorized as follows:
They either translate executable process models into semantically equivalent smart contracts~\cite{corradini2020engineering,klinger2020cross,ladleif2019modeling,lopez2019caterpillar,tran2018lorikeet} or provide interpreter smart contracts which then work on process specifications directly~\cite{lopez2019interpreted,sturm2018lean}, e.g., simulating a token system.
The resulting smart contracts are either used as a connector between different participant's local \glspl{bpms}, or even provide some features of a \gls{bpms} themselves like on\-/chain worklist handlers or monitoring tools.

Regardless the architecture, participants interact with process instances via transactions, during which conditions and constraints specified in the model are checked in the trusted on\-/chain environment.
Referring to the running example (see Fig.~\ref{fig:example}), for instance, the \gls{mno} might be submitting a transaction to the invoicing smart contract to ``add overdue fines'' after they sent a transaction for ``send invoice''.
The smart contract now needs to check whether ``7 days'' have actually passed between the transactions to ensure compliance.

However, determining whether this or other temporal constraints are fulfilled is not trivial, which forms the basis of our problem statement:
Since process instances are smart contracts executed within the confines of the blockchain, the general restrictions mentioned in Sect.~\ref{subsec:blockchain} apply:
\begin{enumerate}
\item Smart contracts are executed in a \textbf{closed\-/world environment} and can not readily access global timing information and clocks.
\item Smart contracts are \textbf{inherently passive} structures only activated within synchronous calls during the mining of transactions and do not offer continuous runtime behavior or timer monitoring.
\item Since transactions originate from single participants and are propagated through a network of mutually distrustful nodes, there is \textbf{no reliable timestamp} attached to them individually, as it could not be independently verified.
\end{enumerate}
These limitations make it hard to pinpoint the exact time a transaction was originally issued, and thus correctly enforce temporal constraints specified in the underlying model.
While the blockchain domain is very diverse and not all technologies might exhibit the same extent of restrictions (see Sect.~\ref{sec:discussion}), existing approaches largely neglect or glance over these issues and the challenges implied~(see Sect.~\ref{sec:related-work}).
In this paper, we contribute a novel and holistic analysis of enforcing temporal constraints found in business processes within blockchain environments, providing possible solutions and evaluating their utility and drawbacks.

\section{Blockchain Formalization}
\label{sec:formalization}
To ground the remaining paper in a formal framework, we abstract the structure of a blockchain to its core components.
This abstraction is based on literature~\cite{xu2017taxonomy} and on implementations like Ethereum~\cite{wood2014ethereum} and Tezos~\cite{goodman2014tezos}, and covers a large variety of blockchains found in practice (see Sect.~\ref{subsec:technological-generality}).

Note that we will not consider low\-/level effects like network latency in our formalization.
We assume that there is an optimal, instantaneous connection between all peers of the blockchain network, and only take into account delays and effects directly related to the blockchain's protocols.

\begin{definition}[blockchain]
\label{def:blockchain}
A blockchain $\mathcal{B} = (B_0, ..., B_n)$ is a cryptographically linked list of $n$ numbered blocks.
Blocks $B_i=(s_i,T_i)$ contain a strictly increasing block timestamp $s_i\in\mathbb{N}$ with $s_{i-1} < s_i \,\forall\, 0 < i \leq n$ and a transaction list $T_i$.
\end{definition}

In literature, the term \emph{block height} is often used interchangeably with block number.
The first block $B_0$ of a blockchain is called \emph{genesis block}.
We regard a timestamp to be a scalar value, for example representing an Unix timestamp.
We do not model the contents of transactions in this paper.
For us, only the exact time a transaction was created by its original sender locally before being propagated within the blockchain network is relevant:

\begin{definition}[transaction timestamp]
\label{def:transaction-timestamp}
For any transaction $tx$, let $s_{tx} \in \mathbb{N}$ be the timestamp the transaction was initially created and signed by its sender.
\end{definition}

\begin{figure}
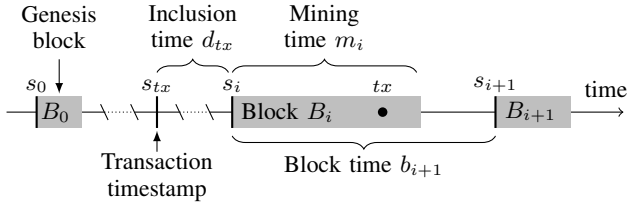

  \centering
  \begin{tikzpicture}
    \input{figures/timeline-width-narrow}
    \input{figures/timeline}
\draw[align=center,decorate,decoration={brace,amplitude=5pt},yshift=2.5*\t] (\tx,0) -- (\bistart-0.05,0) node[anchor=south,midway,above=4pt] {Inclusion\\time $d_{tx}$\vphantom{fg}};

\draw[align=center,decorate,decoration={brace,amplitude=5pt},xshift=0.05,yshift=2.5*\t] (\bistart,0) -- (\biend-0.05,0) node[anchor=south,midway,above=4pt] {Mining\\time $m_i$\vphantom{fg}};

\draw[align=center,decorate,decoration={brace,amplitude=5pt},yshift=-1.5*\t] (\biistart,0) -- (\bistart,0) node[anchor=north,midway,below=4pt] {Block time $b_{i+1}$\vphantom{fg}};

\draw[align=center,latex-] (\bgmid,\t+1pt) -- (\bgmid,2.5*\t+5pt) node[anchor=south,midway,above=7pt] {Genesis\\block};

\draw[align=center,latex-] (\tx,-\t-1pt) -- (\tx,-1.5*\t-6pt) node[anchor=north,midway,below=1pt] {Transaction\\timestamp};
  \end{tikzpicture}
  \caption{Timeline and properties of blocks in a blockchain}
  \label{fig:timeline}
\end{figure}

Figure~\ref{fig:timeline} illustrates an excerpt of a blockchain mapped against a timeline.
The block timestamp $s_i$ of each visible block $B_i$ is fixed at the start of mining, i.e., when a miner bundles the set of transactions and begins performing the mining procedure, which takes some computing time.

\begin{definition}[mining time]
\label{def:mining-time}
Let $\mathcal{B} = (B_0, ..., B_n)$ be a blockchain.
The duration of the mining procedure of a block $B_i$ is called the mining time $m_i$
\end{definition}

Some blockchain networks like Ethereum~\cite{wood2014ethereum} and Bitcoin~\cite{nakamoto2008bitcoin} aim at achieving a fixed block generation rate, and adjust the difficulty of their mining algorithms constantly to react to changes in the network's overall power.
The important measure in this context is block time, the interval between two consecutive blocks (not to be confused with the block timestamp):

\begin{definition}[block time]
\label{def:block-time}
Let $\mathcal{B} = (B_0, ..., B_n)$ be a blockchain.
The block time $b_i$ of a block $B_i$, $0 < i \leq n$, is defined as $b_i := s_i-s_{i-1}$.
\end{definition}

Lastly, after transactions are created and signed by their sender, they need to be propagated, picked up by a miner, and included in a block.
This takes a certain amount of time:

\begin{definition}[inclusion time]
Let $\mathcal{B} = (B_0, ..., B_n)$ be a blockchain and $tx\in T_i$ an arbitrary transaction in some block $B_i=(s_i, T_i)$, $0 < i \leq n$.
Then the delay between the transaction's initial creation $s_{tx}$ and the time it was included in the block is called the inclusion time $d_{tx} := s_i - s_{tx}$ of $tx$.
\end{definition}

\section{Time Measures in Blockchain Systems}
\label{sec:measures}

To correctly execute business processes in smart contracts on the blockchain, the exact time a transaction was originally \emph{created}---the transaction timestamp---needs to be determined, as it eventually decides whether temporal constraints defined in the model were complied with despite potential delays in the blockchain processing.
Since this information is not readily available (see Sect.~\ref{subsec:blockchain}), we propose a number of alternative methods to measure time within smart contracts with different advantages and drawbacks.
Table~\ref{tab:comparison} shows an overview of the measures, and presents a rough relative comparison regarding various metrics on a scale from best to worst.

For simplicity reasons, we will use $tx$ to denote an arbitrary transaction contained in a block $B_i=(s_i, T_i)$, i.e., $tx\in T_i$, of an arbitrary blockchain $\mathcal{B}=(B_0, ..., B_n)$.
The ultimate goal is to acquire a suitable estimate for $s_{tx}$.

\begin{table*}
\newcommand{\xxx}{$\bullet\bullet\bullet$}
\newcommand{\xxo}{$\bullet\bullet\circ$}
\newcommand{\xoo}{$\bullet\circ\circ$}
\newcommand{\ooo}{$\circ\circ\circ$}
\caption{Overview of Time Measures and Relative Comparison on a Scale From Best (\xxx) to Worst (\ooo)}
\label{tab:comparison}
\def\arraystretch{1.1}
\centering
\newcolumntype{Z}{>{\centering\arraybackslash}p{20mm}}
\newcolumntype{Y}{>{\raggedright\scriptsize\arraybackslash}p{20mm}}
\begin{tabular}{l||Z|Z|Z|Z|Z}\hline

\emph{\bfseries Measure} &
\emph{Block timestamp} &
\emph{Block number} &
\emph{Parameter} &
\emph{\Push oracle} &
\emph{\Pull}
\\

&
$\mtimestamp$ &
$\mnumber$ &
$\mparam$ &
$\mpush$ &
\emph{oracle} $\mpull$
\\\cline{2-6}

&
\multicolumn{1}{Y|}{Timestamp included in the block by its miner} &
\multicolumn{1}{Y|}{\scriptsize Number of a block in relation to mean block time} &
\multicolumn{1}{Y|}{\scriptsize Parameter added to a smart contract transaction} &
\multicolumn{1}{Y|}{\scriptsize External provider regularly injecting timestamps} &
\multicolumn{1}{Y}{\scriptsize External provider being queried with separate callback}
\\\hline\hline

\emph{Accuracy} &
\xxo & 
\ooo & 
\xxx & 
\xoo & 
\xoo 
\\\hline

\emph{Trust} &
\xxo & 
\xxx & 
\ooo & 
\xoo & 
\xoo 
\\\hline

\emph{Immediacy} &
\xxx & 
\xxx & 
\xxx & 
\xxx & 
\ooo 
\\\hline

\emph{Cost} &
\xxx & 
\xxx & 
\xxo & 
\ooo & 
\ooo 
\\\hline

\emph{Reliability} &
\xxx & 
\xxx & 
\xxo & 
\xoo & 
\ooo 
\\\hline

\end{tabular}
\end{table*}

\subsection{Definition of Measures}
We differentiate between five measures of time available in the scope of a transaction to a smart contract.
Two of these are tied to the block the transaction is contained in (block timestamp $\mtimestamp$, block number $\mnumber$), one requires manual input (parameter $\mparam$), and two are based on oracles (\push oracle $\mpush$, \pull oracle $\mpull$).

\subsubsection{Block Timestamp}
Transaction $tx$ has access to the timestamp $s_i$ of its including block $B_i$, which can be used as a measure $\mtimestamp(tx) := s_i$.
The miner of $B_i$ fixes the timestamp when they start assembling the block, since it is part of the cryptographically relevant block metadata, and thus crucial in ascertaining the blockchain's integrity.

\subsubsection{Block Number}
The block number of a block $B_i$ can be used to estimate the block timestamp $s_i$, and in extension the timestamps of the transactions contained in it.
That is, if the mean block time is $\bar b$, then $B_i$ would have had to be mined $i$ times $\bar b$ after the genesis block, or more formally at $\mnumber(tx) := s_0 + i \cdot \bar b$.

This measure can not only be used to determine the absolute ``age'' of a block from the genesis block, but also the time interval between two blocks.
For any two arbitrary blocks $B_i$ and $B_j$, they were roughly mined $|i-j|\cdot \bar b$ milliseconds apart.
This is useful for estimating relative time differences, as we will discuss in Sect.~\ref{sec:semantics}.


\subsubsection{Parameter}
Smart contract functions may receive arbitrary arguments encoded in the payload of the calling transaction $tx$.
A function can thus be designed to take in an additional argument, namely a timestamp set by the sender of the transaction, which we will denote with $\mparam(tx)$.
For example, the smart contract function in Listing~\ref{lst:parameter} can only be called if the supplied timestamp is below 1608811200 (noon on Dec. 24\textsuperscript{th}, 2020, UTC).

\lstset{
  frame=single,
  basicstyle=\footnotesize\ttfamily,
  morekeywords={function,uint,require,external},
  morestring=[b]",
  stringstyle=\itshape,
  showstringspaces=false
}
\begin{lstlisting}[
  caption=Solidity Smart Contract Function With Timestamp Parameter,
  label=lst:parameter
]
function withDeadline(uint timestamp) external {
  require(timestamp < 1608811200,
          "deadline missed");
  // ...
}
\end{lstlisting}

\subsubsection{\Push Oracle}
A \push oracle as displayed in Fig.~\ref{fig:push-oracle} could be designed to regularly inject a current trusted real\-/world timestamp into a designated smart contract.
Any $tx$ could then be approximately dated by retrieving this measure $\mpush(tx)$ via an inter\-/contract function call.

\subsubsection{\Pull Oracle}
Time servers are a backbone of distributed computing, and provide clock synchronization through protocols such as the \gls{ntp}.
Such time servers can be used from the context of a blockchain using \pull oracles as shown in Fig.~\ref{fig:pull-oracle} to estimate the transaction timestamp.
To obtain the measure $\mpull(tx)$, a smart contract would schedule a query to a time server, and receive a callback transaction with the measure attached later.

\subsection{Accuracy}
\begin{figure*}
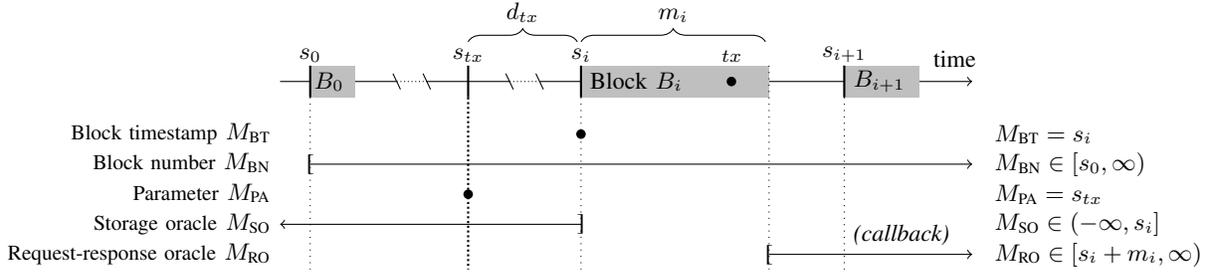

  \centering
  \begin{tikzpicture}
    \input{figures/timeline-width-normal}
    \input{figures/timeline}
\draw[align=center,decorate,decoration={brace,amplitude=5pt},yshift=2.5*\t] (\tx,0) -- (\bistart-0.05,0) node[anchor=south,midway,above=4pt] {$d_{tx}$};

\draw[align=center,decorate,decoration={brace,amplitude=5pt},xshift=0.05,yshift=2.5*\t] (\bistart,0) -- (\biend-0.05,0) node[anchor=south,midway,above=4pt] {$m_i$};
    \input{figures/timeline-accuracy}
    \node[anchor=south west] at (\biistart,\opl) {\itshape (callback)};
    \input{figures/timeline-measures}
  \end{tikzpicture}
  \caption{Accuracy of the measures illustrated by their possible ranges when predicting $s_{tx}$ during the processing of transaction $tx$}
  \label{fig:timeline-measures}
\end{figure*}

The time measures are designed to estimate the transaction timestamp $s_{tx}$, but may deviate considerably in practice.
These deviations are illustrated in Fig.~\ref{fig:timeline-measures}, which displays the possible range of outputs for each measure on a timeline.

The parameter approach $\mparam$ consistently achieves perfect accuracy (assuming honest participants, as we will discuss in the next section) and is the only measure with this property.
Block timestamps $\mtimestamp$ are less accurate in that they always deviate by the inclusion time $d_{tx}$, which depends on several factors like blockchain configuration, congestion, and how much a sender is willing to pay in transaction fees.
However, these factors have been shown to be predictable in principle~\cite{yasaweerasinghelage2017predicting}, meaning a high accuracy can still be achieved.

The \push oracle measure $\mpush$ depends on the frequency the provider updates the timestamp stored in the oracle smart contract.
Since the \push oracle update needs to be performed in the same or an earlier block as $tx$ we can generally assert that $\mpush(tx) \leq s_i = s_{tx} + d_{tx}$, that is, it at most postdates the transaction by its inclusion time.
This is the only guarantee that can be given, as otherwise the measure could be arbitrarily outdated.
The \pull oracle $\mpull$ behaves rather obversely and at least postdates $s_{tx}$ by $d_{tx}+m_i$ if the oracle provider immediately reacts upon observing the request in $B_i$.
Since this can not be guaranteed and the oracle provider may only react later, the measure can be arbitrarily inaccurate.

For the block number, fluctuations in the block times might lead to arbitrary discrepancies in both directions.
As an illustrative example, we considered the block times of the Ethereum main chain observed within the last five years\footnote{Data from 2015-07-30--2020-03-18, \url{https://etherscan.io/chart/blocktime}} and arrived at a mean $\bar b$ of 15.19\,s (min. 4.46\,s, max. 30.31\,s) with a standard deviation of 2.71\,s.
Extrapolated linearly from the genesis block, this would for example result in a timestamp of 2020-03-29T06:58:49Z for block 9690267, when actually it was mined at 2020-03-17T16:55:27Z, some 12 days earlier.

It should be noted at this point that time itself is not an entirely static concept.
Being regulated by international and national groups, minor or major adjustments may occasionally be prescribed which can not be accommodated for in all systems.
As a concrete example, leap seconds are irregularly added to \gls{utc}, but not reflected in Unix timestamps as used by Ethereum~\cite{wood2014ethereum}.
Thus, the block measures can not take leap seconds into account, in contrast to the parameter and oracle approaches.

\subsection{Trust}
Blockchains are designed to be trustless in that no participant, be it a miner or a business partner, can deviate from agreed\-/upon specifications, and the integrity of the data can be independently verified.
However, some degree of trust in other entities is required for some time measures.

The block number approach $\mnumber$ relies only on the inherent blockchain protocol and configuration, and cannot be unilaterally forged since block numbers must strictly increase by one each block.
Miners thus can not arbitrarily choose block numbers without violating the integrity constraint, making this the measure with the least trust assumptions.

The block timestamp measure $\mtimestamp$ on the other hand is set by the miner, and there is a degree of freedom in fixing it.
Ethereum, for example, demands a ``reasonable''~\cite{wood2014ethereum} timestamp, and Tezos similarly postulates that the ``protocol design must tolerate reasonable clock drift''~\cite{goodman2014tezos}.
Bitcoin even amended its protocol in 2016 to fix a monetary ``incentive for miners to lie about the time of their blocks''\footnote{Bitcoin Improvement Proposal (BIP) 113, \url{https://git.io/JfYqt}, 2020-04-29}.
However, consensus algorithms in practice do constraint the window allowed by this:
For one, younger blocks will usually be preferred since they were mined first.
Secondly, some implementations place an upper limit on the deviation, e.g., 15 seconds in the future when receiving a new block for Ethereum\footnote{Ethereum Ethash protocol (Go version), \url{https://git.io/JvdNc}, 2020-04-02}.

For oracles, trust lies entirely in the third\-/party oracle provider to serve correct information.
Oracle providers have acknowledged this issue and offer various mitigation options.
For example, Provable uses proof systems to certify that they did not tamper with the data provided from external APIs\footnote{\url{http://docs.provable.xyz/#security-deep-dive}, 2020-04-20}.
Other oracles rely on second\-/layer consensus algorithms themselves to decrease trust~\cite{lo2020reliability}.
Combined with the monetary incentives to attract and keep customers, we assess the required trust to be on a par with the block timestamp.

Finally, the parameter approach performs worst in this metric, since trust is placed in each collaborator in the smart contract to attach correct timestamps $\mparam$ to the transactions.
Acknowledging that the outspoken lack of trust between participants is one of the main reasons to use blockchain technology in the first place~\cite{weber2016untrusted}, this severely reduces the measure's utility.

\subsection{Immediacy}
The time measure should be immediately available to the smart contract code when it is executed via a transaction.
This is the case for all approaches except the \pull oracle, which provides the measure via a delayed callback transaction.
The time measure can thus not be used in\-/place, but there needs to be some kind of waiting mechanism which increases the complexity of the smart contracts involved.

\subsection{Cost}
Cost is crucial in determining the practical use of a measure.
In general, blockchain transactions incur a cost, the so\-/called \emph{transaction cost}, which is distributed to the miners to incentivize their operations.
Transaction cost usually depends on both the size of the transaction's payload, as well as on the computational effort required to execute the called smart contract function.

We base our comparison on whether a measure incurs any significant extra cost on top of the regular cost of the transaction.
This is not the case for the block timestamp $\mtimestamp$ or block number $\mnumber$, which are available ``for free'' in the context of the transaction by default.
The parameter measure $\mparam$ requires adding additional payload, which will increase the overall cost marginally.
The oracle measures $\mpush$ and $\mpull$ require an inter\-/contract function call, which is typically quite expensive.
Additionally, the providers of the oracles will probably incur some kind of fee for their service.
For example, Provable prices regular API calls at USD 0.01, up to USD 0.04 if a notary proof is requested\footnote{\url{http://docs.provable.xyz/#pricing}, 2020-04-20}.

\subsection{Reliability}
A measure should be reliably accessible, i.e., be resilient to attacks and system failures.
This is the case for both the block timestamp $\mtimestamp$ as well as the block number $\mnumber$ since they are an inherent part of the blockchain protocol, and necessarily available whenever a transaction is executed.
The parameter measure $\mparam$ is slightly less reliable as it depends on participants attaching a timestamp to each transaction.
Mistakes in calculating the timestamp and sending the transaction could thus lead to deviations or errors.

For the oracle measures $\mpush$ and $\mpull$, reliability varies since a third\-/party service provider is involved~\cite{lo2020reliability}.
This issue is less severe in \push oracles since a dedicated smart contract with some return value will always be available---albeit with potentially outdated or uninitialized data---even if the provider experiences issues.
\Pull oracles would not be able to return a value, however, if the provider is offline.
Especially long\-/running processes could get stuck if the provider used when the smart contract was initially deployed is not available anymore after some months or years.

\section{Application to Process Execution}
\label{sec:semantics}

In this section, we will evaluate whether the time measures are suitable for the implementation of temporal constraints found in business processes; in particular absolute timers, relative timers, cycles, as well as deferred choice (see Sect.~\ref{subsec:time-in-bpm}).

\subsection{Absolute Timers}
Absolute temporal constraints model deadline scenarios, in which certain activities have to start or finish by a specific point in time.
The start event in the running example (see Fig.~\ref{fig:example}), for instance, governs that the process may not start before the 1\textsuperscript{st} of any specific month.
In terms of blockchain\-/based process execution, that means that any transaction $tx$ corresponding to either the timer start event itself or the ``send invoice'' activity may not be accepted by the smart contract before that.
If we assume that $s_e$ is the scalar timestamp of midnight on the 1\textsuperscript{st} of some month, this comes down to determining whether $s_e \leq s_{tx}$ within the smart contract.

\begin{figure*}
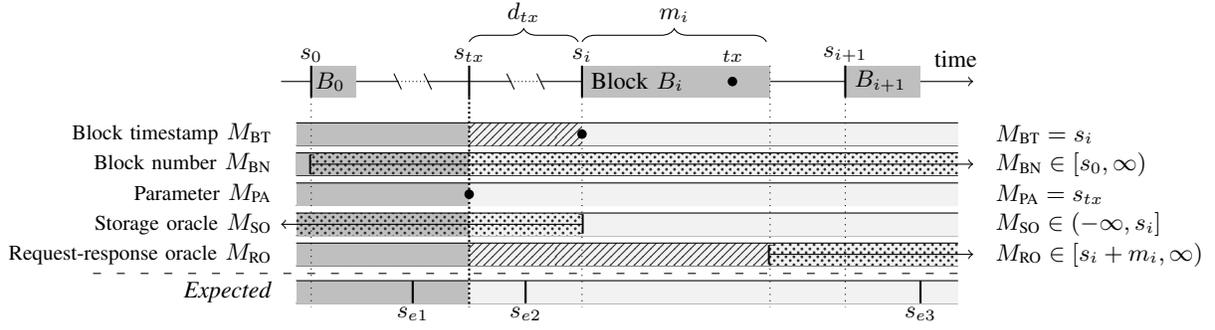

  \centering
  \begin{tikzpicture}
    \input{figures/timeline-width-normal}
    \input{figures/timeline}
\draw[align=center,decorate,decoration={brace,amplitude=5pt},yshift=2.5*\t] (\tx,0) -- (\bistart-0.05,0) node[anchor=south,midway,above=4pt] {$d_{tx}$};

\draw[align=center,decorate,decoration={brace,amplitude=5pt},xshift=0.05,yshift=2.5*\t] (\bistart,0) -- (\biend-0.05,0) node[anchor=south,midway,above=4pt] {$m_i$};
    \input{figures/timeline-absolute}
    \input{figures/timeline-accuracy}
    \input{figures/timeline-measures}
  \end{tikzpicture}
  \caption{Possible occurence of false negatives/positives (dotted pattern) and permanent false positives (diagonal line pattern) for absolute temporal constraints}
  \label{fig:timeline-absolute}
\end{figure*}

For any time measure $M$, one of two fundamental errors may be observed:
a \emph{false negative} occurs if $M(tx) < s_e \leq s_{tx}$, that is, the measure erroneously reports that $s_e$ has not happened yet although it did;
and a \emph{false positive} occurs if $s_{tx} < s_e \leq M(tx)$, that is, the measure erroneously reports that $s_e$ has elapsed even when it did not.

Figure~\ref{fig:timeline-absolute} shows whether the time measures exhibit those errors.
The expected result is that elapsed deadlines relative to $s_{tx}$, like $s_{e1}$, are correctly recognized as past (dark background) and prospective ones like $s_{e2}$ or $s_{e3}$ as still open (light background).
A dotted pattern on a measure's timeline signifies that false positives or negatives are possible in the interval, while a diagonal line pattern more severely signifies that the false result will \emph{always} be obtained in the interval.

For example, consider $s_{e2}$ which happens after the transaction timestamp $s_{tx}$, but before it is included in a block, and should receive a result indicating it is still open.
The block timestamp measure $\mtimestamp(tx) = s_i$ will always falsely report the deadline as past.
The \push oracle measure $\mpush(tx)$ might yield any timestamp in the time interval up to $s_i$, thus potentially leading to a false positive.
Similar errors happen for the \pull oracle and block number measures.
Only the parameter approach $\mparam$ is consistently correct.

\begin{table}
\newcommand{\xxx}{$\bullet\bullet\bullet$}
\newcommand{\xxo}{$\bullet\bullet\circ$}
\newcommand{\xoo}{$\bullet\circ\circ$}
\newcommand{\ooo}{$\circ\circ\circ$}
\caption{Relative Comparison of Time Measures When Implementing Temporal Constraints From Best (\xxx) to Worst (\ooo)}
\label{tab:constraints}
\def\arraystretch{1.1}
\centering
\newcolumntype{Z}{>{\centering\arraybackslash}p{15mm}}
\newcolumntype{Y}{>{\raggedright\scriptsize\arraybackslash}p{15mm}}
\begin{tabular}{l||Z|Z}\hline

\emph{\bfseries Temporal constraint} &
\emph{Absolute} &
\emph{Relative}
\\\hline\hline

\emph{Block timestamp} $\mtimestamp$ &
\xxo &
\xxo
\\\hline

\emph{Block number} $\mnumber$ &
\ooo &
\xoo
\\\hline

\emph{Parameter} $\mparam$ &
\xxx &
\xxx
\\\hline

\emph{\Push oracle} $\mpush$ &
\xoo &
\ooo
\\\hline

\emph{\Pull oracle} $\mpull$ &
\xoo &
\ooo
\\\hline

\end{tabular}
\end{table}

Table~\ref{tab:constraints} summarizes these findings and ranks the measures on a scale from best to worst.
The parameter approach $\mparam$ objectively performs best when absolute temporal constraints are concerned, and always yields the right result.
The block timestamp $\mtimestamp$ will result in a false positive if the event trigger happens between the creation of the transaction $tx$ and its inclusion in a block, making it slightly worse, but still relatively predictable~\cite{yasaweerasinghelage2017predicting}.

The other measures, however, may perform arbitrarily bad.
The \push oracle $\mpush$ may return any timestamp before $s_i$, and thus is only perfectly accurate for later deadlines.
The \pull oracle $\mpull$ may provide an arbitrary timestamp after the propagation of block $B_i$ at $s_i+m_i$, and thus the measure is only accurate for elapsed deadlines.
Consequently, the oracle measures only seem sensible when strictly trying to avoid false positives (\push oracle) or false negatives (\pull oracle).

On the lower end of the spectrum, the block number $\mnumber$ is only technically accurate if the deadline elapsed before the timestamp $s_0$ of the genesis block, i.e., before the blockchain was even started, since that is the lower bound of timestamps obtained per definition.
Otherwise, results are arbitrary.

\subsection{Relative Timers}
\begin{figure}
  \centering
  \begin{tikzpicture}
    \input{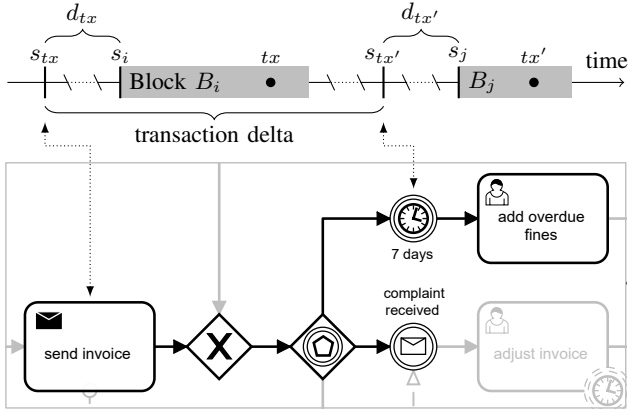}
  \end{tikzpicture}
  \caption{Relative temporal constraints as deltas between transactions}
  \label{fig:timeline-relative}
\end{figure}

Relative temporal constraints concern minimum or maximum delays between different activities.
In the running example in Fig.~\ref{fig:example}, for instance, the ``7 days'' until overdue fines are added start once the invoice has been sent and the following event\-/based gateway has been enabled.
Supporting relative temporal constraints thus comes down to calculating the time difference or \emph{delta} between the two corresponding transactions.
This is illustrated in Fig.~\ref{fig:timeline-relative}, in which an excerpt of a blockchain is drawn alongside the corresponding snippet from the running example.
The issue is determining in $tx'$ whether 7 days have passed since $tx$ was created.

Table~\ref{tab:constraints} summarizes the findings regarding the time measures' suitability in resolving these kinds of relative temporal constraints.
As is the case for the absolute case, only the parameter approach $\mparam$ consistently achieves perfect results, that is, $\mparam(tx')-\mparam(tx)=s_{tx'}-s_{tx}$
The block timestamp $\mtimestamp$ performs predictably worse due to its inherent deviations from the transaction timestamp

The block number measure $\mnumber$ performs reasonably well if we employ the alternative time interval definition given in Sect.~\ref{sec:measures}.
That is, we do not calculate $\mnumber(tx')-\mnumber(tx)$, but directly consider $|i-j|\cdot\bar b$, the total block times between $B_i$ and $B_j$.
Because of the smaller multiplicator, the inaccuracy due to fluctuations in the block time is reduced.

In general, though, the inaccuracy of a measure multiplies when using it to calculate deltas, which becomes especially apparent for the oracle measures.
While the \push oracle measure $\mpush$ might be adequate when trying to minimize false positives on absolute temporal constraints, the difference between two timestamps obtained from a \push oracle at different times might be arbitrarily large if there are no further requirements on the service which is queried.
The same is true for \pull oracles.

\subsection{Cycles}
Cycles are complex temporal constraints in that they can start at an absolute or a relative point in time, and combine the difficulties discussed in both the previous sections.
Thus, from a temporal perspective, cycles are covered by our discussion of absolute and relative temporal constraints, respectively.

The main issue is making sure that all iterations of the cycle are hit.
For example, the process in the running example (see Fig.~\ref{fig:example}) needs to be started at the beginning of each month, and no month may be missed.
This leads to implementation overhead in the smart contracts, which have to keep track of some kind of counter to make sure this is not the case.

\subsection{Deferred Choice}
The deferred choice pattern introduces a major complication, which is the lack of causality between the events to compare.
Whereas for relative temporal constraints, we know that we have to compare the delta between two transactions which must always arrive in the same order, as governed by the sequence flow in the model, deferred choice means that any transaction could arrive at any time.

Consider the process snippet from the running example in Fig.~\ref{fig:timeline-relative}, for instance.
The event\-/based gateway models a deferred choice between ``7 days'', ``complaint received'' and ``payment received'' (clipped in the figure).
Per the semantics of \gls{bpmn}, the transaction first created should be accepted.
There is no way for the smart contract to decide on this, however:
If the customer has complained already and the transaction for ``complaint received'' is pending, it could be overtaken by the ``7 days'' transaction even if it was actually sent in time.
Overdue fines could then be added mistakenly.

The underlying cause for this is that it is the miner's decision which transactions to include in a block from the pending pool, and how to order them.
Miners are not aware of domain and application\-/specific constraints such as deferred choice, and decide based on purely economical or protocol\-/related reasons.
As such, we can not give a clear indication as to the performance of the different time measures.

\section{Results and Evaluation}
\label{sec:discussion}

In this section, we summarize our results on the running example, deriving more general application guidelines, and evaluate the overall approach in broader terms.

\subsection{Results and Guidelines}
In Sects.~\ref{sec:measures} and \ref{sec:semantics}, we compared the time measures regarding their inherent properties as well as their utility in implementing temporal constraints, respectively.
A holistic result, of course, requires more than just naively ``summing up the points''---which would yield that the block timestamp $\mtimestamp$ and parameter approach $\mparam$ heavily outperform all other measures, in particular those resorting to oracles.

Instead, we propose more nuanced results.
While we can confidently attest that temporal constraints can be implemented in blockchain\-/based process execution, the exact degree to which this is the case entirely depends on the scenario and the requirements of all involved participants.
The comparisons provided in this paper as a whole may be taken as a guideline when evaluating measures for a concrete use case to help arriving at a meaningful choice.

Take, for instance, the running example from Fig.~\ref{fig:example}, in which an \gls{mno} and its customer handle an invoice.
At first glance, the parameter approach $\mparam$ seems like a good choice, since it provides perfect accuracy, eliminating legally questionable situations in which fines are added prematurely.
However, the customer may have doubts about the credibility of those timestamps, as they are usually not in a perceived position of power in face of a large corporation like an \gls{mno}.
In turn, the \gls{mno} might take care of thousands of customers, making them relucant to place trust in customer\-/provided timestamps which are hard to oversee.
Thus, the block timestamp $\mtimestamp$ seems like the more sensible choice, since the relative deficit in accuracy rarely comes into play when looking at durations of days or months.
In practice, deviations due to processing times---like for bank transfers---are well\-/understood and should pose no further problems when applied to blockchain technology.

\subsection{General Evaluation}
The fields of \gls{bpm} and blockchain technology are highly complex and diverse in their own rights, let alone in their interplay.
In this section, we will briefly evaluate whether this has an impact on the results of this paper.

\subsubsection{Technological Generality}
\label{subsec:technological-generality}
The blockchain model we introduced in Sect.~\ref{sec:formalization} is based on common blockchains such as Ethereum~\cite{wood2014ethereum} and Tezos~\cite{goodman2014tezos}, but adopts a somewhat traditional structure for the blockchain and its protocols which initially lead to the restrictions identified in Sect.~\ref{subsec:blockchain}.
There are blockchains which work differently, though, and may not experience the same difficulties:

Hyperledger Fabric~\cite{androulaki2018hyperledger} uses sophisticated network structures and provides endorsement\-/based transaction timestamps.
Corda R3~\cite{hearn2019corda} features a concept called ``time\-/windows'', in which notaries enforce certain temporal constraints.
Ostensibly, having a participant, notary, or endorser provide a timestamp is equivalent to the parameter approach $\mparam$ introduced in this paper in many aspects---trust is merely shifted.
Nevertheless, novel blockchains might open up new perspectives and issues in the future.

\subsubsection{Conceptual Completeness}
We did not consider some blockchain concepts like forks.
As a consequence confirmation time, i.e., the time a participant has to wait in practice until they can be reasonably sure a transaction was successfully included in the blockchain, is not taken into account.
This does not majorly impact our results, though.
From a smart contract's perspective, forks are not visible and thus do not affect any time measure.
From a participant's perspective, personal risk tolerance is highly individual, and confirmation times might vary greatly~\cite{yasaweerasinghelage2017predicting}.

\subsubsection{Coverage of Temporal Constraints}
\label{subsec:coverage}
While we have largely covered inter\-/activity and inter\-/process temporal constraints in this paper, we did not dive deeper into intra\-/activity constraints like activity durations~\cite{cheikhrouhou2015temporal}.
These kinds of constraints are rarely found in blockchain\-/based business process execution approaches since transactions usually have a very brief time limit to avoid deadlocks and ensure throughput.
This inherently limits the time a transaction can spend processing an activity.
Nevertheless, intra\-/activity temporal constraints have been shown to be in principle feasible within single blockchain transactions~\cite{abid2020modelling}.

Conceivably, constraints on task durations could also be implemented using at least two transactions for each activity, one at the start and one at the end of working on it.
Additionally, task duration constraints in \gls{bpmn} have been shown to be expressible via inter\-/activity temporal constraints using specific patterns of timer and signal events~\cite{combi2017modeling}.
This would make them susceptible to the same techniques we used in this paper for the other temporal constraints.

\section{Related Work}
\label{sec:related-work}
Temporal constraints are a well\-/researched area in \gls{bpm}~\cite{eder1999time,cheikhrouhou2015temporal}.
Yet, little research has been done regarding the implementation of temporal constraints in blockchain\-/based process execution.
Notable frameworks like Caterpillar~\cite{lopez2019caterpillar}, Lorikeet~\cite{tran2018lorikeet}, and others~\cite{corradini2020engineering,lopez2019interpreted,sturm2018lean} do not mention them at all.
Some approaches briefly acknowledge some challenges involved and leave them for future work~\cite{ladleif2019modeling,weber2016untrusted}, or seem to support some notion of temporal constraints without discussing the actual implementation~\cite{klinger2020cross}.
Some exceptions do exist, however.

Abid et al. extend Caterpillar to support the modeling and execution of several intra and inter\-/activity temporal constraints like task durations and absolute start/end times~\cite{abid2020modelling}.
Their implementation on Ethereum uses time\-/guards checking block timestamps---called $\mtimestamp$ in this paper---to evaluate those constraints, but the authors do not provide a critical discussion of the consequences or feasibility of that decision.
Similar restrictions apply to Mavridou and Laszka, who use finite state machines as their source modeling language and support timed transitions within them, again implemented using block timestamps on Ethereum~\cite{mavridou2017designing}.

Lastly, work has been carried out regarding the simulation and prediction of temporal aspects in blockchain\-/based process execution.
Yasaweerasinghelage et al. predict more technical properties of blockchain networks like confirmation times and how they impact business process execution~\cite{yasaweerasinghelage2017predicting}.
Haarmann annotates choreography models with activity\-/level duration information and estimates the total duration of the choreography in face of varying inclusion times and block times through simulation~\cite{haarmann2019estimating}.
Both works abstract from the concrete source of the timing information, i.e., they assume there is some measure readily available.

In summary, we assert that no approach provides a systematic discussion and evaluation of temporal constraints and their implementation on blockchain.
Important metrics like trust or accuracy, alternative measures like oracle\-/based timestamps, or more advanced patterns of constraints like deferred choice are not considered.
Thus, to the best of our knowledge, this paper contributes the first holistic framework of its kind.

\section{Conclusion and Future Work}
\label{sec:conclusion}
Temporal constraints are of utmost importance in business processes.
When implementing compliant blockchain\-/based smart contracts for these processes, however, the blockchain environment's inherent restrictions pose several challenges.
In this paper, we have identified these challenges, and proposed a number of alternative measures facilitating temporal constraints.
We have compared these measures, and provided hints as to their utility in process execution.
In particular, the results and their critical assessment can be used as a guideline to choose the right measures for specific scenarios.
We hope that our contributions may help in enabling the development of more sophisticated and complete blockchain\-/based business process execution platforms.

In future work, we aim to explore further techniques not discussed in this paper.
Second\-/layer consensus protocols could be used to let participants in a process dispute timestamps they suspect are invalid.
Heartbeat transactions could trigger the smart contract regularly to globally check enabled constraints, and reduce the potential for false positives or negatives later on.
Lastly, social and organizational aspects surely warrant a deeper look, for example regarding the actual level of trust between collaborating enterprises and which trade\-/offs in enforcing temporal constraints are acceptable in practice.
This could help to further broaden the acceptance of blockchain technology in \gls{bpm} in the near future.

\section*{Acknowledgment}
We would like to thank Ingo Weber for his valuable input and feedback on early versions of the manuscript.

\IEEEtriggeratref{11}
\bibliographystyle{IEEEtran}
\bibliography{bibliography}

\end{document}